# Temperature-Aware Virtual Data Center Embedding to Avoid Hot Spots in Data Centers

Chao Guo, Kai Xu, Gangxiang Shen, *Senior Member, IEEE*, and Moshe Zukerman, *Fellow, IEEE*

*Abstract*—Virtual Data Center (VDC) embedding has drawn significant attention recently because of growing need for efficient and flexible means of Data Center (DC) resource allocation. Existing studies on VDC embedding mainly focus on improving DCs' resource utilization. However, an important problem that has not been considered in VDC embedding solutions is the creation of hot spots by excessive heat dissipation and hot air generation from racks in DCs, which have significant adverse effect on energy consumption of the cooling system and IT equipment lifespan. To address this issue, we propose a temperature-aware VDC embedding scheme to avoid hot spots by minimizing the maximum temperature of hot air emitted from each rack. Meanwhile, we also aim to reduce the total power consumption of IT equipment in this scheme. A Mixed Integer Linear Programming (MILP) model and a heuristic algorithm are developed to implement the proposed VDC embedding scheme. Numerical results show that the proposed temperature-aware embedding scheme can significantly outperforms a load-balanced embedding scheme in terms of maximum rack temperature, total power consumption of IT equipment, and VDC rejection ratio.

## I. INTRODUCTION

SERVER virtualization technologies (e.g., VMWare [1] and Xen [2]) have been widely adopted in today's Data Centers (DCs). Through such virtualization, computing and storage resources in each physical server are sliced into multiple Virtual Machines (VMs) of different sizes. Each of the VMs can be viewed as a standalone machine running different operational systems and applications, which therefore improves efficiency and flexibility of server resources allocation. However, this is still far from ideal because current Data Centers (DCs) do not provide performance isolation for network resources, without which network performance degradation may adversely influence performance of deployed cloud services [3]. Fortunately, emerging technologies, such as Network Function Virtualization (NFV) [4] and Software Defined Networking (SDN) [5], provide solutions that address these problems. Based on these techniques, new DC

Part of this work was conducted when the first author was with Soochow University. This work was supported by a grant from National Key R&D Program China (2018YFB1801701) and a grant from the Research Grants Council of the Hong Kong Special Administrative Region, P.R. China (CityU 11200318). (The corresponding author: Gangxiang Shen)
C. Guo was with Soochow University before he joined the Department of Electrical Engineering, City University of Hong Kong, Hong Kong, SAR, P.R. China; C. Guo, K. Xu, and M. Zukerman are with the Department of Electrical Engineering, City University of Hong Kong, Hong Kong, SAR, P.R. China.
G. Shen is with the School of Electronic and Information Engineering, Soochow University, Suzhou, Jiangsu, P.R. China. (email: shengx@suda.edu.cn)

virtualization architectures, such as SecondNet [6], NetLord [7], have been proposed to fully virtualize resources in a DC. In addition to provision VMs only, in these virtualization architectures, DCs provide resources and services in terms of Virtual Data Centers (VDCs). Each VDC is composed of a set of VMs connected by a set of virtual links. Each virtual link is assigned with a certain amount of guaranteed bandwidth which provides isolation of communication resources and guarantees Quality of Service (QoS) of communications [3].

VDC is a promising cloud service paradigm, where Service Providers (SPs) lease resources from Infrastructure Providers (InPs) to deploy their own applications, and then provide services to their end users [3]. Amazon EC2 [8] and Microsoft Azure [9] are typical InPs. They own hyper-scale Data Centers (DCs) and can provide vast amounts of computing, storage, and network resources. Under the VDC service provisioning paradigm, SPs do not need to build their own physical DCs, which can significantly reduce their capital expenditure (CAPEX) and operational expenditure (OPEX).

An important and challenging problem for data center virtualization is VDC embedding, which addresses the issue of allocating resources to SPs through mapping VDCs onto physical nodes and links in a DC. Different approaches have been proposed to solve this problem, many of which are to maximize resource utilization, reduce energy consumption and system cost [10-14]. However, in the context of VDC embedding, there is still an important problem that have not been considered, namely, hot spots created by excessive heat dissipation and hot air generation from racks can significantly impact the efficiency of a cooling system and lifespans of IT equipment in a DC. We need to address the issue of heat generation in a DC when embedding VDCs.

In today's DCs, servers and Top of Rack (ToR) switches are often densely mounted on racks for space saving [3][15]. In such circumstances, hot air emitted from these devices forms a hot spot at the back of the rack, causing serious problems, including the risk of equipment failure [16][17]. It was reported that the failure rate of a physical server operated at 40°C is 66% higher than 20°C [18]. To avoid hot spots, DCs are usually equipped with a dedicated cooling system which unfortunately consumes significant energy. It was reported that about 50% of total power consumption in a DC is from the cooling system, much more than the power consumed by IT equipment (i.e., servers and network elements like switches) [19]. What is worse, the efficiency of a cooling system decreases dramatically when the supplied cooling temperature drops [20]. In the context of VDC embedding, when local temperature at a certain location is too high due to improper VDC embedding, the cooling system needs to lower the supplied temperature, which will lead to a sharp increase of additional power consumption by the cooling system [20]. Therefore, when embedding VDCs, it is important to consider also the heat generation issue.

Different from existing studies, for the first time, we consider the problem of heat generation in the context of VDC embedding. We propose a temperature-aware VDC embedding scheme that avoids hot spots in a DC by appropriately embedding VDCs onto different physical equipment. The main contributions and the key novelty of this study are summarized as follows.

- *We propose a temperature-aware VDC embedding scheme to avoid local hot spots*. In this scheme, we try to avoid local hot spots by minimizing the maximum temperature of each rack when embedding VDCs. In addition, we also try to minimize the total energy consumption of the IT equipment in a DC when embedding VDCs.

- *We provide a Mixed Integer Linear Programming (MILP) model for the problem of temperature-aware VDC embedding*. The MILP model enables us to find an optimal solution for the static VDC embedding scenario, where all the VDC requests are known in advance. For this, we assume that physical resources in a DC are sufficient to accommodate all the VDC requests. The objective of the MILP model is to jointly minimize the maximum temperature of each rack, and the total power consumption of all IT equipment in a DC.

- *We develop a temperature-aware VDC embedding algorithm*. The heuristic algorithm aims to achieve the same goal as the MILP model. As the algorithm has a lower computational complexity, it is efficient to embed VDCs under a large DC scenario. Moreover, it can also be employed to embed VDCs under dynamic VDC embedding scenario, where resources in a DC is limited and some VDC requests may be rejected due to lack of computing or communication resources. For this scenario, we also set a threshold for rack temperature. A VDC request will be rejected if the temperature of a rack is higher than the threshold after embedding the VDC.

- *We evaluate the proposed temperature-aware embedding scheme for both static and dynamic scenarios*. For the static scenario, we consider both the MILP model and the heuristic algorithm. Results show that the heuristic algorithm can achieve performance close to the MILP model. Moreover, the results also show that the proposed embedding scheme can significantly outperform the load-balanced scheme in terms of the maximum rack temperature and the total power consumption of all IT equipment. For the dynamic scenario, we evaluate the VDC rejection ratio (defined as the total number of rejected VDC requests divided by the total number of arrived VDC requests), and results show that, the proposed temperature-aware embedding scheme has significantly lower rejection ratios under different temperature thresholds than the load-balanced scheme.

The remainder of this paper is organized as follows. In Section II, we summarize related works on VDC embedding. In Section III, we introduce the architecture of a DC and the basic idea of the temperature-aware VDC embedding scheme. Section IV presents our research problem and the corresponding MILP model. Section V introduces the proposed heuristic algorithm for temperature-aware VDC embedding. We evaluate the performance of both the MILP model and the heuristic algorithm in Section VI. Section VII concludes the paper.

## II. RELATED WORK

### A. VDC Embedding

VDC embedding has drawn significant attention in recent years. Related studies have mainly focused on the following aspects.

First, energy consumption is always an important concern for InPs, and therefore, many studies have focused on *energy-efficient* VDC embedding. Nam *et al*. [10][11] proposed to minimize energy consumption of each VDC by using the fewest physical resources to accommodate the VDC and further reduce energy consumption for dynamic VDC requests by applying VM migration and server consolidation. Similarly, Zhani *et al.* [12] proposed a dynamic VDC embedding scheme, called VDC Planner, which employs VM migration to maximize revenue, and minimize scheduling delay and total energy consumption of the whole IT system. Yang *et al.* [21] introduced virtual switches for a VDC and proposed two different approaches to embed VDCs with the objective of minimizing energy consumption. In [13] and [14], Amokrane *et al*. studied the problem of VDC embedding for a geographically distributed DC and considered to reduce carbon footprint through reducing energy consumption and utilizing renewable energy. Han *et al*. [22] also considered the problem of VDC embedding with the objective of reducing energy consumption of both computing and communications in a distributed DC environment. In addition, in [44], we proposed mixed VDC embedding that is capable of supporting both unicast and multicast services and developed a mixed embedding scheme to minimize system cost and energy consumption.

Second, as an important concern for cloud users, how to enhance *availability* and *reliability* of VDCs has also been extensively studied. Zhang *et al*. [23] proved that computing of VDC availability is NP-hard and proposed a heuristic algorithm to compute VDC availability. They also proposed a framework, called Venice, to ensure high availability for embedded VDCs. Wen *et al*. [24] considered trade-off between reliability, revenue, and bandwidth occupation when embedding VDCs, and proposed new embedding algorithms to improve the revenue of an InP. In [25], Sun *et al*. studied the problem of reliable VDC embedding in a geographically distributed DC environment. Lo *et al*. [26] proposed a framework, called CALM, to avoid adverse effects caused by switch failures. Yu *et al*. [27] also considered the problem of reliable VDC embedding and proposed an algorithm to jointly optimize primary and backup VDC embedding.

The third important research challenge in the area of VDC embedding is how to reduce the rejection ratio of VDCs when provisioning VDC services so that InPs can increase revenue and meet users' QoS requirements. Gilesh *et al*. [28][29] proposed to minimize the rejection ratio through VM migration and resource defragmentation when embedding VDCs. Yan *et al.* [30] considered network congestion and proposed an algorithm to release traffic loads on some busy links through selectively relocating certain VMs. By doing this, they improve the VDC acceptance ratio. Also, Sun *et al*. [31] reduced the VDC rejection ratio by migrating VMs among different DCs.

In addition to the above research directions, some researchers also considered other research aspects in the field of VDC embedding, such as the latency [32] and security issues [33]. Also, in our previous work [45], we looked into the problem of performance degradation due to load burstiness of embedded VDCs. However, none of the existing studies on VDC embedding design and optimization have focused on the problem of heat generation in DCs.

*B. Cooling-Aware Resource Allocation in DCs*

Heat generated by physical equipment severely impacts their reliability. Many studies have considered this issue when allocating resources in DCs. Moore *et al.* [20] noticed that cooling cost in a DC is significant and therefore proposed system-level methods to reduce the cooling cost through temperature-aware workload placement. Similarly, Tang *et al*. [34] studied the chassis-level heat recirculation problem where hot air emitted from a chassis may return back and increase the temperature of air drawn into the chassis. They proposed an algorithm to improve cooling efficiency and save cooling cost. In [35], Li *et al*. proposed a holistic algorithm to schedule VMs in a DC to minimize total energy consumption by both cooling system and physical servers, while ensuring the CPU temperature of each server not to violate a threshold. In [16], Choi considered the rack-level heat generation problem when studying VM placement and proposed a heuristic algorithm which allocates different number of backup servers according to different rack temperatures. In [36], Ilager *et al.* considered eliminating hot spots through properly scheduling VMs in a DC.

There are still many publications focusing on the problem of heat generation in a DC, e.g., [37] and [38]. However, all studies that consider the heat generation problem have not considered VDC embedding.

*C. Summary*

In summary, though there are many existing studies on VDC embedding, they have not considered the problem of heat generation in a DC. Also, though there are many publications related to the problem of heat generation in a DC, they are not within the context of VDC embedding. Therefore, to the best of our knowledge, the current work is the first to consider the problem of heat generation when embedding VDCs.

III. PROBLEM OF HEAT GENERATION IN VDC EMBEDDING

In this section, we first introduce the architecture of a DC, the layout of IT equipment in a DC, and the concept of VDC embedding. Following this, we present the model of heat generation employed in this study and further illustrate the proposed temperature-aware VDC embedding.

*A. VDC Embedding in a DC*

A DC houses many different types of IT equipment, such as physical servers and network elements (e.g., switches, routers, and communication links). They provide computing, storage, and communication resources. To maintain a constant temperature in a DC (so as to avoid equipment failures due to local overheating), a DC also hosts a dedicated cooling system. In addition, a dedicated energy system is required to power all the IT equipment and the cooling system in a DC [3].

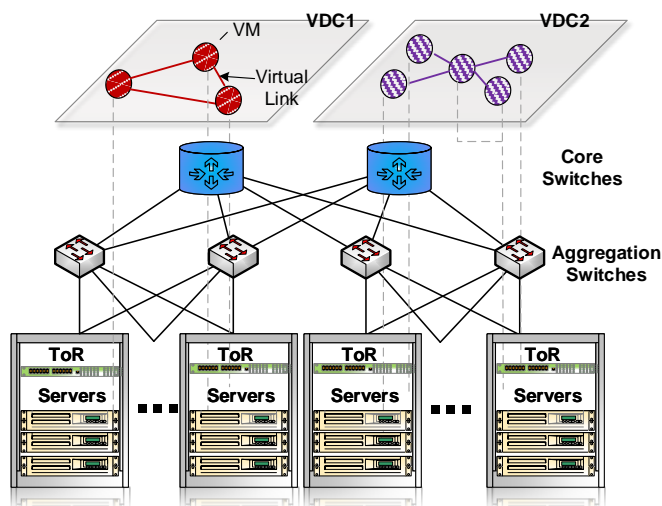

Fig. 1. DCN architecture and VDC embedding.

DC Network (DCN) connects physical servers and network elements in a DC following a certain topology.

Fig. 1 illustrates a DCN, where physical servers are densely packed in racks to save space, and each server in a rack is directly connected to a physical switch on top of the rack (ToR). Such a switch is called ToR switch. ToR switches are further connected to higher-level switches to form a tree like topology in three tiers. In Fig. 1, the bottom tier is made up of ToR switches, each of which is connected to several aggregation switches in the middle tier. Each aggregation switch is responsible for aggregating traffic from several ToR switches and is further connected to several core switches in the top tier. Different from ToR switches, aggregation and core switches are not mounted on racks. Instead, they are placed in a so-called "central physical area" in a DC [15].

VDC embedding is to provide DC resources to SPs through mapping virtual elements required by a VDC onto physical IT equipment in a DC. In Fig. 1, there are two VDCs, i.e., VDC1 and VDC2. Each VDC consists of several VMs and virtual links, which form a certain virtual topology. Each VM requires certain amounts of computing and storage resources, and it is mapped onto a physical server. Each virtual link requires a certain amount of network bandwidth, which is mapped onto a physical path traversing multiple physical links and switches.

*B. Heat Generation Model*

Before introducing temperature-aware VDC embedding, we first introduce the model of heat generation adopted in this study. We consider rack-level heat generation. The heat generated by a rack has a direct relationship with the total power consumption of IT equipment in the rack, given as [16][34]:

$$T_{out} = T_{in} + \frac{P_{rack}}{\rho f c_p}, \qquad (1)$$

where $T_{out}$ is the *outlet temperature*, that is the temperature of hot air generated from the rack, $T_{in}$ is the *inlet temperature*, that is the temperature of cold air drawn into the rack, and $P_{rack}$ is the total power consumption of IT equipment in the rack. $\rho$, $f$, and $c_p$ are thermo-physical values, where $\rho$ [kg/m³] is the density of air, $f$ [m³/s] is the airflow rate in the rack, and $c_p$ is the specific heat capacity of the air, i.e., the amount of energy that one unit mass of air needs to absorb or release when the temperature rises or decreases by one degree, in units of [kJ/(kg·°C)].

To avoid hot spots, we minimize the outlet temperature of each rack. According to (1), we observe that the outlet temperature of each rack is linear to both inlet temperature and total power consumption of the rack. In addition, the power consumption of physical equipment, such as a server, linearly increases with its workload [34]. Therefore, an intuitive approach to minimize the outlet temperature of each rack is to place more workload on colder racks while less workload on hotter racks. By doing this, IT equipment located at hotter racks will generate less heat and the outlet temperature can be lowered according to (1). In addition, to further reduce the power consumption of each rack, we use the smallest number of servers and physical links in the rack to support the required workload. This is because IT equipment consumes additional power to keep itself active in addition to being linear to its workload [34].

It should be noted that it takes some time to stabilize the outlet temperature of a rack when its workload changes. However, for simplicity, in this study we assume that the time of this stabilization is negligible. Also, similar to [16], we assume that the hot air emitted from a rack does not return back to the inlet of the rack.

*C. Example of Temperature-Aware VDC Embedding*

We use the example in Fig. 2 to illustrate the basic idea of temperature-aware VDC embedding. Suppose that there are two VDC requests, i.e., VDC0 and VDC1. VDC0 consists of three VMs, i.e., VMs 0, 1, and 2, and three virtual links. VDC1 consists of two VMs, i.e., VM3 and VM4, and one virtual link. Assume that each VM has the same size and each physical server can hold two such VMs. Also, we assume that the bandwidth of each physical link in the physical DCN is ample to establish the virtual links of the two VDCs. In addition, to guarantee reliability, we require VMs from a single VDC to be mapped onto different servers.

If not considering the issue of heat generation, we may adopt the energy-efficient embedding scheme shown in Fig. 2(b). In this scheme, to minimize total energy consumption, all VMs are mapped onto the fewest servers in the same rack, i.e., servers in rack R0 in Fig. 2(b), and virtual links are mapped onto the fewest physical links in the same rack. Because the two VDCs are mapped onto R0 simultaneously, R0 has a high workload, and since R0 has a high inlet temperature because it is located far away from the cooling source (e.g., 21°C), the outlet temperature is very high (e.g., 29°C), forming a hot spot. In contrast, because R1 has no workload, it has lower inlet and outlet temperatures, forming a cold spot. Though the scheme can minimize the energy consumption of IT equipment, for the whole DC, which also hosts a cooling system to maintain a constant environment temperature, the energy-efficient scheme is not efficient because the DC consumes more energy in the cooling system so as to avoid the hot spot.

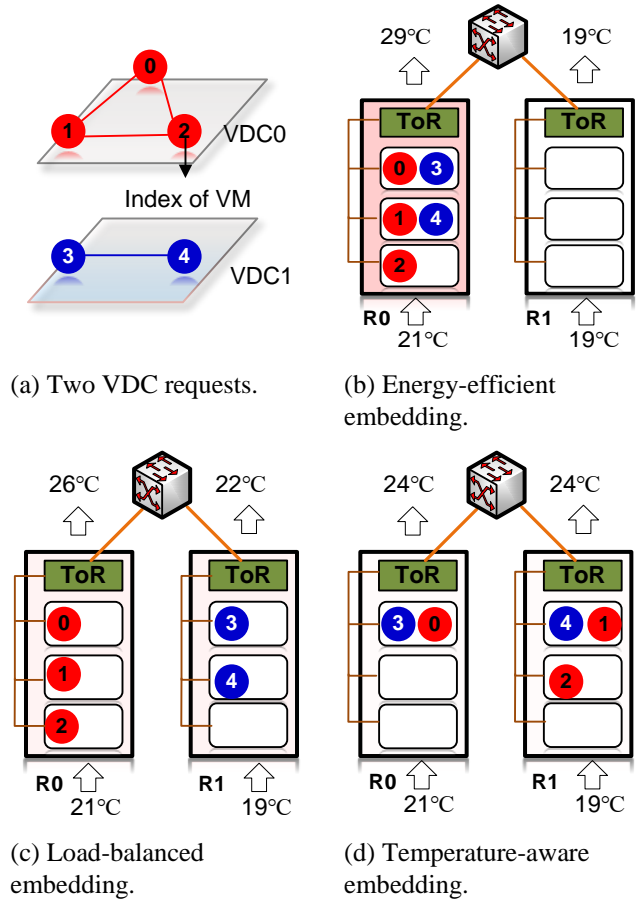

(a) Two VDC requests.  (b) Energy-efficient embedding.

(c) Load-balanced embedding.  (d) Temperature-aware embedding.

Fig. 2. Embedding VDCs based on three different schemes.

Alternately, we may also consider a load-balanced scheme as illustrated in Fig. 2(c), where the two VDCs are placed in two different racks. This scheme helps balance the workload on each rack, but still may not be optimal in balancing rack temperatures. This is because the scheme ignores the temperature information, and therefore, may assign more workload to R0, which leads to a higher outlet temperature. Another drawback of this scheme is that, in both R0 and R1, each server holds one VM, which requires more servers to be active. As a result, the total power consumption of each rack is still not minimized.

Finally, Fig. 2(d) illustrates our temperature-aware scheme. Different form the other two schemes, the proposed scheme maps more VMs onto R1 that has a lower inlet temperature. Moreover, in each rack, VMs are consolidated onto fewer servers such that the power consumption of the rack is minimized. In Fig. 2(d), the outlet temperature of the two racks can be balanced to 24°C, much lower than those of the other two schemes. Moreover, compared with the load-balanced scheme, the temperature-aware scheme also reduces the total energy consumption of all IT equipment because fewer servers are used.

## IV. Temperature-Aware VDC Embedding Problem and MILP Model

In this section, we present the temperature-aware VDC embedding problem and then formulate this problem into an MILP optimization model.

### A. Problem Statement

The problem of temperature-aware VDC embedding is as follows. We are given a physical DCN which consists of a set of servers, switches, and physical links. Each server provides certain amounts of computing and storage resources. Each physical link provides a certain amount of network bandwidth. We are also given a set of physical racks, and each rack hosts a set of servers and a ToR switch. The inlet temperature of each rack is assumed to be known in advance.

Also, we assume that there is a set of VDC requests. Each VDC consists of a set of VM nodes and a set of virtual links interconnecting these VM nodes. Each VM requires certain amounts of computing and storage resources. Each virtual link requires a certain amount of bandwidth. We consider two types of VDC request scenarios, namely, static and dynamic scenarios.

In the static scenario, all the VDC requests are given in advance, and we assume that resources in the DC are enough to accommodate all the requests. The objective is to minimize the maximum outlet temperature of all the racks and the total power consumption of IT equipment. We formulate the problem into an MILP model and also propose a scalable heuristic algorithm to solve the problem.

In the dynamic scenario, VDC requests arrive randomly and sequentially. At the end of their service times, these VDCs will leave the DC. In this scenario, we consider to minimize their rejection ratio. Also, to avoid hot spots, we set an outlet temperature threshold, and if the threshold is violated after embedding a VDC, we reject this request.

### B. MILP Model

The MILP model for solving the static VDC embedding problem is formulated as follows.

**Sets:**

| | |
|---|---|
| $\Phi$ | Set of racks. |
| $N$ | Set of server nodes. |
| $S$ | Set of switch nodes, including ToR, aggregation, and core switches. |
| $NE_n$ | Set of neighboring nodes of node $n$. Here $n$ can be a server node or a switch node. |
| $R$ | Set of server resource types, including CPU, memory, and disk space. |
| $\Omega$ | Set of VDC requests |
| $V_i$ | Set of VM nodes in VDC $i$ |

**Parameters:**

| | |
|---|---|
| $\beta_{nk}$ | A binary parameter to denote that whether server or switch node $n$ is placed in rack $k$. |
| $C_{rn}$ | Capacity of resource $r$ in server $n$. |
| $B_{mn}$ | Capacity of physical link $(m, n)$, where $m$ and $n$ are the two end nodes of the link. |
| $D_r^{iv}$ | Amount of resource $r$ required by VM $v$ of VDC $i$. |
| $\mu_{sd}^i$ | Bandwidth required by virtual link $(s, d)$ of VDC $i$. Here, $s$ and $d$ are the two end VM nodes. |
| $P_{PM}^{idle}$ | Power consumption of a server when it is idle. |
| $P_{switch}^{idle}$ | Power consumption of a switch when it is idle. |
| $P_{PM}^{max}$ | Power consumption of a server when it is fully loaded. |
| $P_e, P_o$ | Power consumption of an electronic and an optical switch port, respectively. |
| $T_k^{in}$ | Inlet temperature of rack $k$. |
| $\rho, f, c_p$ | Three thermo-parameters, which are density of air, airflow rate, and specific heat capacity of air, |

respectively.

Δ  A large number.

α  A factor weighting the optimization of maximum rack outlet temperature and the total power consumption of IT equipment.

**Variables:**

$\delta_{vn}^i$  A binary variable that takes the value of one if VM $v$ of VDC $i$ is mapped onto server $n$; zero, otherwise.

$\mu_{mn}^{i,sd}$  A real variable that denotes the amount of bandwidth reserved for virtual link $(s,d)$ on physical link $(m,n)$.

$\omega_n$  A binary variable that takes the value of one if server or switch $n$ is active; 0, otherwise.

$\pi_{mn}$  A binary variable that takes the value of one if physical link $(m,n)$ is used; zero, otherwise.

$q_n, \sigma_n$  Two integer variables that denotes the number of electronic and optical switch ports used on switch $n$, respectively.

$P_n$  A real variable that denotes the power consumption of server or physical switch $n$.

$T_k^{out}$  Outlet temperature of rack $k$.

$T$  Maximum outlet temperature of all racks.

The objective of the MILP model is to minimize the maximum rack outlet temperature and the total power consumption of all IT equipment while accommodating all the VDC requests.

**Objective**: minimize

$$T + \alpha \cdot \sum_{n \in N \cup S} P_n. \tag{2}$$

To achieve the objective, the following constraints must be satisfied.

**Constraints**:

- **VM Mapping Constraints**

$$\sum_{n \in N} \delta_{vn}^i = 1 \ \forall i \in \Omega, v \in V_i \tag{3}$$

$$\sum_{v \in V_i} \delta_{vn}^i \leq 1 \ \forall i \in \Omega, n \in N \tag{4}$$

$$\sum_{i \in \Omega, v \in V_i} \delta_{vn}^i \cdot D_r^{iv} \leq C_{rn} \ \forall r \in R, n \in N. \tag{5}$$

- **Virtual Link Mapping Constraints**

$$\sum_{m \in NE_n} \mu_{nm}^{i,sd} - \sum_{m \in NE_n} \mu_{mn}^{i,sd} = (\delta_{sn}^i - \delta_{dn}^i) \cdot \mu_{sd}^i \ \forall i \in \Omega, s, d \in V_i: s \neq d, n \in N \tag{6}$$

$$\sum_{m \in NE_n} \mu_{nm}^{i,sd} - \sum_{m \in NE_n} \mu_{mn}^{i,sd} = 0 \ \forall i \in \Omega, s, d \in V_i: s \neq d, n \in S \tag{7}$$

$$\mu_{mn}^{i,sd} = \mu_{nm}^{i,ds} \ \forall i \in \Omega, s, d \in V_i: s \neq d, m \in N \cup S, n \in NE_m \tag{8}$$

$$\sum_{i \in \Omega, s, d \in V_i: s \neq d} \mu_{mn}^{i,sd} \leq B_{mn} \ \forall m \in N \cup S, n \in NE_m. \tag{9}$$

- **Active Server and Switch Constraints**

$$\omega_n \geq \delta_{vn}^i \ \forall i \in \Omega, v \in V_i, n \in N \tag{10}$$

$$\omega_n \leq \sum_{i \in \Omega, v \in V_i} \delta_{vn}^i \ \forall n \in N \tag{11}$$

$$\Delta \cdot \pi_{mn} \geq \mu_{mn}^{i,sd} \ \forall i \in \Omega, s, d \in V_i: s \neq d, m \in N \cup S, n \in NE_m \tag{12}$$

$$\Delta \cdot \pi_{mn} \geq \mu_{nm}^{i,sd} \ \forall i \in \Omega, s, d \in V_i: s \neq d, m \in N \cup S, n \in NE_m \tag{13}$$

$$\pi_{mn} \leq \sum_{i \in \Omega, s, d \in V_i: s \neq d} (\mu_{nm}^{i,sd} + \mu_{mn}^{i,sd}) \ \forall m \in N \cup S, n \in NE_m \tag{14}$$

$$\omega_m \geq \pi_{nm} \ \forall m \in S, n \in NE_m \tag{15}$$

$$\omega_m \geq \pi_{mn} \ \forall m \in S, n \in NE_m. \tag{16}$$

- **Numbers of Electronic and Optical Switch Ports Used in Each Physical Switch**

$$q_n = \sum_{m \in NE_n \cap N} \pi_{nm} \ \forall n \in S \tag{17}$$

$$\sigma_n = \sum_{m \in NE_n \cap S} \pi_{nm} \ \forall n \in S. \tag{18}$$

- **Power Consumption by Server and Physical Switch**

$$P_n = \omega_n \cdot P_{PM}^{idle} + (P^{max} - P_{PM}^{idle}) \cdot \sum_{i \in \Omega, v \in V_i, r \in "CPU"} \frac{\delta_{vn}^i \cdot D_r^{iv}}{C_{rn}} \ \forall n \in N \tag{19}$$

$$P_n = \omega_n \cdot P_{switch}^{idle} + q_n \cdot P_e + \sigma_n \cdot P_o \ \forall n \in S. \tag{20}$$

- **Outlet Temperature Constraints**

$$T_k^{out} = T_k^{in} + \frac{1}{\rho f c_p} \cdot \sum_{n \in N \cup S} \beta_{nk} \cdot P_n \quad \forall k \in \Phi \tag{21}$$

$$T \geq T_k^{out} \quad \forall k \in \Phi. \tag{22}$$

**VM Mapping Constraints:** Constraint (3) ensures that each VM is successfully mapped onto a single physical server. Constraint (4) ensures that a server can host no more than one VM from the same VDC $i$. We consider one to one VM mapping in this paper, where different VMs from a common VDC are required to be mapped onto different servers. Constraint (5) ensures the resource restriction of each server.

**Virtual Link Mapping Constraints:** Constraints (6) and (7) are the flow conservation equations to build up a physical path for virtual link $(s, d)$. Constraint (8) ensures that the physical path for each virtual link is bidirectional. Constraint (9) ensures the resource restriction of each physical link.

**Active Server and Switch Constraints**: Constraints (10) and (11) ensure that if there is a VM mapped onto a server, this server must be active. Constraints (12)-(14) decide whether a physical link is used. Constraints (15) and (16) ensure that if one of the physical links connected to a switch is used, this switch must be active.

**Numbers of Electronic and Optical Switch Ports Used in Each Physical Switch**: Constraints (17) and (18) find the numbers of electronic and optical switch ports used. We assume that electronic switch ports are used to connect servers and ToR switches, while optical switch ports are used to provide communication capability between physical switches.

**Power Consumption by Server and Physical Switch:** Constraint (19) calculates the power consumption of a server. We adopt the model of server power consumption proposed in [34], which is linear to the CPU utilization plus an idle power. Constraint (20) calculates the power consumption of a physical switch, which sums its idle power and the power consumption of electronic and optical switch ports.

**Outlet Temperature Constraints:** Constraint (21) calculates the outlet temperature of each rack, which follows the equation of heat generation model (1). Constraint (22) finds the maximum outlet temperature of all racks.

16The computational complexity of an MILP model is dominated by the numbers of variables and constraints. In this model, the dominant number of variables is at the level of $O(|\mathbf{\Omega}| \cdot |\mathbf{V}|^2 \cdot |\mathbf{N}| \cdot |\mathbf{NE}|)$ due to variable $\mu_{mn}^{i,sd}$, where $|\mathbf{\Omega}|$ is the total number of VDC requests, $|\mathbf{V}|$ is the total number of VMs in each set $\mathbf{V}_i$ ($i \in \mathbf{\Omega}$), $|\mathbf{N}|$ is the total number of physical nodes, and $|\mathbf{NE}|$ is the total number of neighboring nodes in each set $\mathbf{NE}_m$ ($m \in \mathbf{P} \cup \mathbf{S}$). The dominant number of constraints is also at the level of $O(|\mathbf{\Omega}| \cdot |\mathbf{V}|^2 \cdot |\mathbf{N}| \cdot |\mathbf{NE}|)$ due to constraint (8).

## V. Heuristic Algorithm for Temperature-Aware VDC Embedding

Although the MILP solution is optimal, it is not suitable for a realistic scale of DCN due to high computational complexity. Therefore, we propose a scalable heuristic algorithm which aims to achieve the same goal as the MILP optimization model while having a much lower computational complexity. Alg. 1 presents the pseudocode of the heuristic algorithm, which consists of two main steps: VM mapping and virtual link mapping. We detail these two steps as follows.

| Alg. 1: Temperature-Aware VDC embedding |
|---|
| 1.  **If (MapVMs() ==**True) //VM Mapping |
| 2.     Allocate resources for VMs; |
| 3.  **Else** |
| 4.     Withdraw resources allocated to the VMs; |
| 5.     **Return False;** //VM mapping fails, return. |
| 6.  Sort virtual links according to bandwidth demands in descending order; |
| 7.  **For** ($j = 1; j \leq m; j++$) //$m$ is the number of virtual links |
| 8.     **If** (**FindPath()** == True) //**Virtual Link Mapping** |
| 9.     Allocate resources for the $j^{th}$ virtual link; |
| 10.    Else |
| 11.     Withdraw resources allocated to the VDC; |
| 12.     **Return False;** //Virtual link mapping fails, return. |
| 13.  **Return True.** //VDC embedding succeeds, return. |

**VM Mapping**: This step corresponds to *MapVMs*() procedure in Alg. 1. Fig. 3 illustrates the detail of VM mapping. The left-hand part of Fig. 3 shows the flowchart of the step and the right-hand part shows the example to illustrate each step in the flowchart. In the VM mapping step, we first sort all the VMs in a VDC according to its CPU demands in descending order. And for each of the VMs in the sorted list, we map it as follows. First, we try to map the VM onto a server held in the coldest rack (i.e., the rack with minimum outlet temperature). For this, we first sort racks according to their outlet temperatures in ascending order, and then scan the racks in the ordered list to find the first eligible rack as a candidate rack. Then, in the candidate rack, we further sort servers according to their remaining CPUs in ascending order, and find the first one that has sufficient CPU, memory, disk, and port bandwidth to host the VM. Doing this can help use the fewest inactive servers. Finally, we map the VM onto this server.

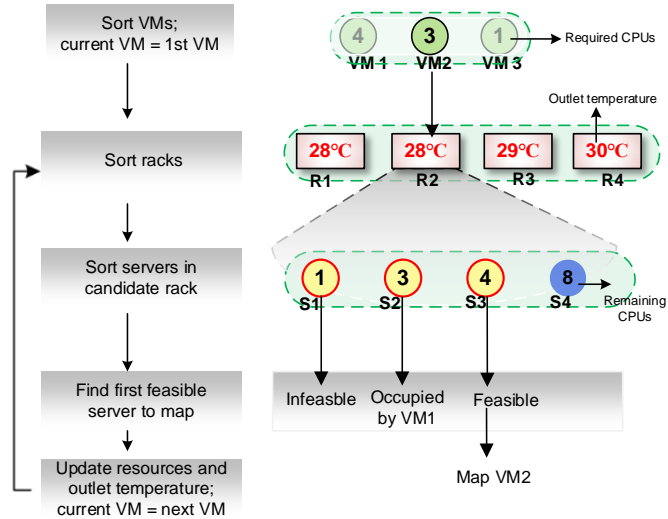

Fig. 3. Flowchart and exmaple of VM mapping procedure.

In the example of VM mapping (see the right-hand part of Fig. 3), before mapping VM2, we first sort racks R1-R4 according to their outlet temperatures in ascending order. If no servers in rack R1 can host VM2, we consider rack R2 as a candidate rack. Once we find a candidate rack, we then sort servers S1-S4 according to their remaining CPU resources in ascending order and find that S2 is the first server with sufficient resources for hosting VM2. However, because S2 has been occupied by VM1, it is not an eligible server because of the reliability requirement, i.e., two VMs belonging to a common VDC should not be embedded in the same server. Therefore, S3 is selected as an eligible server and VM2 is mapped onto it.

After successfully mapping a VM onto a server, we update the load on the server and the outlet temperature of its hosted rack. For the remaining VMs, we repeat the same steps until all the VMs are mapped.

**Virtual Link Mapping**: This step corresponds to the *FindPath*() procedure in Alg. 1. After finishing VM mapping, we map virtual links one by one in descending order of their required bandwidths. To set up a virtual link $l_j$, we search for a path in the physical DCN as follows.

We first remove all links from the physical DCN, each of which the remaining bandwidth is smaller than the bandwidth required by $l_j$. For the remaining physical links in the DCN, we set their costs as their capacity utilizations. If the load on a link is zero, we set its cost to be a large value (because we prefer to use physical links that have already been used), so as to minimize the number of switch ports used to further reduce power consumption. Based on the remaining topology, we search the shortest path between two server nodes that host the two end VMs of $l_j$. We then set up the virtual link along the physical path by reserving corresponding bandwidth on each physical link traversed by the path. We repeat the above steps until all virtual links are successfully mapped.

We next analyze the computational complexity of Alg. 1 as follows. Alg. 1 includes a sorting process multiple times. Because different sorting processes may have different computational complexities [39], to simplify analyses, we assume the complexity of each sorting process used in Alg. 1 is $T(n)$, where $n$ is the number of elements to be sorted. Alg. 1 consists of two main steps, i.e., VM mapping and virtual link mapping. The complexity of VM mapping is $O(T(|V|) + |V| \cdot T(|\Phi|) + |V| \cdot |\Phi| \cdot T(|P|))$, where $|V|$ is the number of VMs in a VDC, $|\Phi|$ is the number of racks in a DC, and $|P|$ is the number of servers in each rack. The complexity of virtual link mapping is $O(T(|VL|) + |VL| \cdot |N|^2)$, where $|VL|$ is the total number of virtual links in the VDC and $|N|$ is the total number of physical nodes in the physical DCN. Therefore, the overall computational complexity of Alg. 1 is $O(T(|V|) + |V| \cdot T(|\Phi|) + |V| \cdot |\Phi| \cdot T(|P|) + T(|VL|) + |VL| \cdot |N|^2)$.

The above embedding algorithm is used to embed one VDC. To embed multiple VDCs, we need to consider two scenarios, i.e., static and dynamic VDC embedding. Next, we introduce these two scenarios.

*A. Static Scenario*

For the static scenario, all the VDC requests are given in advance and the resources in a DC is assumed to be sufficient for accommodating all the requests. In this scenario, we first sort all the VDC requests according to their numbers of VMs in descending order and then apply Alg. 1 to map these VDCs one by one. After embedding all the VDCs, we calculate the maximum outlet temperature of all the racks and the total power consumption of all IT equipment.

*B. Dynamic Scenario*

For the dynamic scenario, because resources in a DC is limited, some requests may be rejected when the remaining resources is not enough. To judge whether a VDC request should be rejected, we set a threshold for the outlet temperature and this threshold should not be violated so as to avoid hot spots. Specifically, for an incoming VDC, we first try to embed it using Alg. 1, and if the remaining resources are not enough for accommodating the VDC, or if the outlet temperature of some racks violet the threshold after embedding the VDC, we reject it. In this scenario, we also consider turning off physical equipment if they are idle so as to further reduce power consumption and outlet temperatures. We count the number of rejected VDCs after a certain number of requests have arrived (e.g., one million) and then calculate the rejection ratio, which is defined as the total number of rejected VDCs divided by the total number of arrived VDCs.

## VI. PERFORMANCE EVALUATION

In this section, we evaluate the performance of the proposed temperature-aware VDC embedding scheme. We consider both the static and dynamic VDC request scenarios. For the static scenario, we evaluate the performance of the scheme based on both the MILP model and the heuristic algorithm. We consider the maximum outlet temperature of racks and the total power consumption of all IT equipment. For the dynamic scenario, we evaluate the VDC request rejection ratio of the scheme.

*A. Experimental Conditions*

We employ the VL2 topology [40] as our DCN and consider two test cases. The first case (i.e., Case A) consists of 20 servers, 4 ToR, 2 aggregation, and 2 core switches. We consider 4 racks and each rack hosts 5 servers and 1 ToR switch. The other test case (i.e., Case B) consists of 400 servers, 40 ToR, 20 aggregation, and 10 core switches. The number of racks in this case is set to be 40 and each rack hosts 10 servers and 1 ToR switch. The link between each server and the ToR switch in the same rack is a twisted copper wire at 1-Gb/s data rate. Each link between two switches is an optical fiber at 10-Gb/s data rate. Each server is assumed to have 100 CPU units, 1,000 memory units, and 10,000 units of disk space. The power consumption of each server is set to be 0.2 kW when it is idle, and 0.5 kW when it is fully loaded. The power consumption of each switch is set to be 0.04 kW when it is idle, and the power consumption of each active electronic and optical switch port is set to be 0.01 kW and 0.08 kW, respectively. The inlet temperature of each rack is randomly (uniformly) generated within [15, 20]°C in the evaluation study. The three thermo-physical values in formula (1) are set the same as those in [16] and [34]. Specifically, the air density is set to be 1.19 kg/m$^3$, the air flow rate in each rack is set to be 0.2454 m$^3$/s, and the specific heat capacity of the air is set to be 1.005 kJ/(kg·°C).

The number of VMs per VDC is set to vary from $M = 2$ to 6 in Case A, and $M = 2$ to 12 in Case B. In both cases, CPU requirement per VM varies from 5 to 30 units, memory requirement per VM varies from 0 to 100 units, disk space requirement per VM varies from 0 to 1,000 units, and bandwidth requirement per virtual link varies from 10 to 70 Mb/s. We first generate all the VMs in a VDC, and then use the following procedure to generate the virtual topology. We start from an empty graph and randomly select a VM node to add to the graph. Then, we execute the following loops until all the remaining VMs are added to the graph. In the $i^{th}$ loop, we first randomly select a VM node (denoted as $VM_i$) that has not been added to the graph. Then, we randomly select a number (denoted as $N_{vm}$) of VM nodes already in the graph and generate a virtual link between $VM_i$ and each of these $N_{vm}$ VM nodes. $N_{vm}$ is randomly (uniformly) generated within $[1, E]$, where $E$ is the number of VM nodes already in the graph. Finally, we add $VM_i$ and all the new virtual links to the graph.

We also consider the *load-balanced* scheme for performance comparison. For this, we adopt the least load algorithm for VM mapping where each VM is mapped onto a server with the lowest CPU load [41], and we apply the Least Load Routing (LLR) algorithm [42][43] for virtual link mapping.

We employed the commercial software package AMPL/Gurobi to solve the MILP model, and the weight factor $\alpha$ in the objective function (2) is set to be 0.1 by considering the minimization of maximum outlet temperature as the first priority. The MIPGAP was set to be 0.01%. We used Java to implement the heuristic algorithms.

## B. Results of Static Scenario

### 1) Maximum Rack Outlet Temperature

We first evaluate the maximum outlet temperature of racks subject to the condition that all VDCs are embedded. Fig. 4 shows the results of the different schemes, where "MILP" corresponds to the MILP model, "Alg_TA" corresponds to the temperature-aware embedding algorithm, and "Alg_LB" corresponds to the load-balanced embedding algorithm.

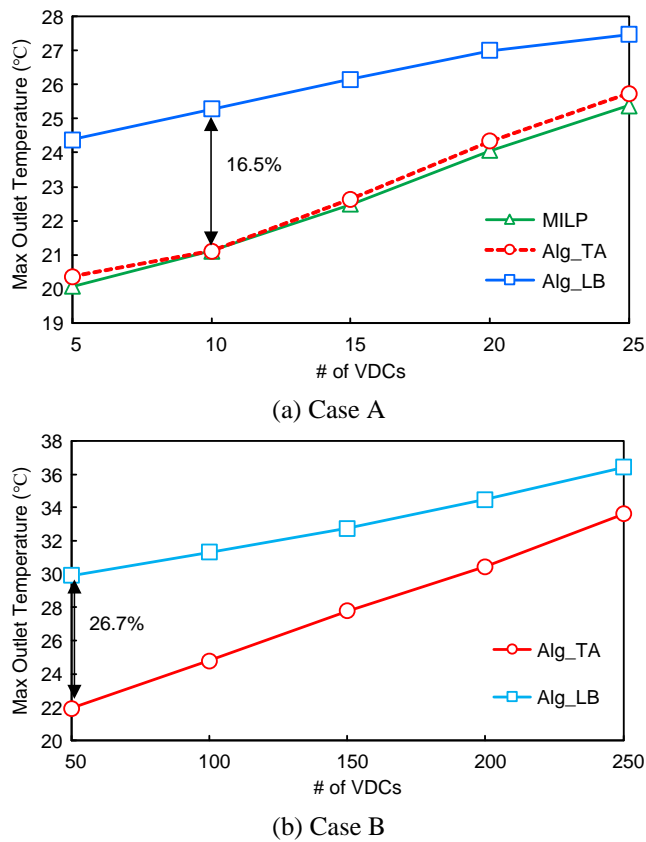

(a) Case A

(b) Case B

Fig. 4. Maximum outlet temperature of racks vs. increasing number of VDCs.

Fig. 4(a) shows the results of Case A. From the results, we observe that the temperature-aware embedding scheme can significantly lower the maximum outlet temperature compared with the load-balanced scheme, up to 16.5%. This is because the temperature-aware scheme maps each VM onto a server in the coldest rack. Moreover, in each rack, the temperature-aware scheme uses the fewest servers and switch ports, which helps further reduce the outlet temperature of the rack and the total power consumption in each rack. In contrast, the load-balanced scheme does not consider the temperature information, which leads to much higher power consumption in each rack. Therefore, its performance is worse than that of the temperature-aware scheme. However, we see that the difference between the two schemes decreases as the number of VDCs increases. This is because as the workload increases, more servers and switch ports are used (become active) in each rack, and therefore the remaining optimization benefit gradually decreases. Also, we observe that the performance of our proposed temperature-aware algorithm is very close to that of the MILP model, which further verifies the efficiency of the former.

Similar studies were carried out for Case B, of which results are shown in Fig. 4(b). Here we do not provide the results of the MILP model as it is computationally intractable to solve the problem due to its large size. Again, we see that the temperature-aware scheme performs much better than the load-balanced scheme, and the reduction of maximum rack outlet temperature is more than 26%, which again verifies the effectiveness of the temperature-aware scheme in avoiding hot spots in a DC.

*2) Distribution of Rack Outlet Temperatures*

To further evaluate the performance of the different embedding schemes, we also analyze the distribution of outlet temperatures of all racks. We consider Case B for this analysis, in which the number of VDCs is 200.

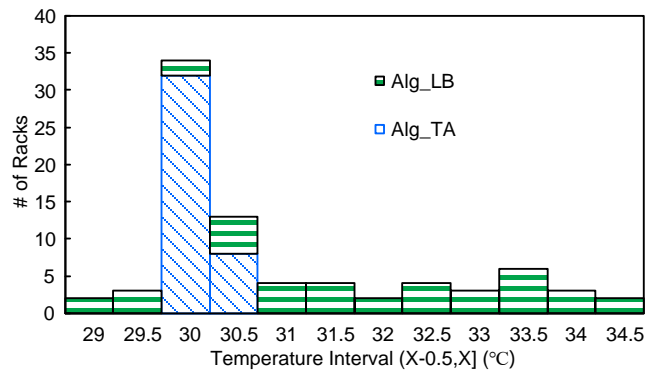

Fig. 5. Number of racks vs. rack outlet temperature intervals.

Fig. 5 shows the numbers of racks in different outlet temperature intervals. If the outlet temperature of a rack is within a certain temperature interval, then we count this rack for this temperature interval. From the results, we observe that, under the temperature-aware scheme, the outlet temperatures of all racks are distributed within the interval (29.5, 30.5]°C, which corresponds to only 1°C difference. This means that the temperature-aware scheme can well balance the temperature distribution of all racks and therefore avoid the advent of hot spots. In contrast, under the load-balanced scheme, the outlet temperatures are evenly distributed within the interval (28.5, 34.5]°C, which spans 6°C, that is much larger than that of the temperature-aware scheme. This means that, under the load-balanced scheme, some racks are cold while others are quite hot. Moreover, most of the racks under the load-balanced scheme are hotter than those under the temperature-aware scheme. Therefore, from the perspective of total power consumption, this difference shows that the temperature-aware scheme is more energy-efficient in running the cooling system than the load-balanced scheme.

*3) Total Power Consumption of IT Equipment*

In addition to the outlet temperature, we also evaluate the total power consumption of all IT equipment in a DCN, including the total power consumption of servers and switches. The results are shown in Fig. 6. Fig. 6(a) shows the results of Case A. From the results, we see that the temperature-aware scheme has a much lower (more than 43%) total power consumption of IT equipment than the load-balanced scheme. This is because the load-balanced scheme assigns workloads evenly onto physical equipment, which requires many servers and switch ports to be used. As a result, the total power consumption of the load-balanced scheme is much higher. In contrast, the temperature-aware scheme uses the fewest servers in each rack and the fewest physical links when embedding VDCs, which helps minimize the number of servers and switches used, so the power consumption can be significantly reduced compared to the load-balanced scheme. Moreover, we see that the results of the temperature-aware algorithm are very close to the optimal results obtained by the MILP model, further verifying the effectiveness of the proposed scheme.

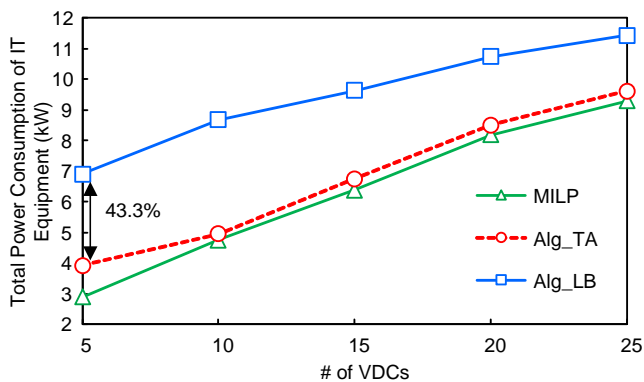

(a) Case A

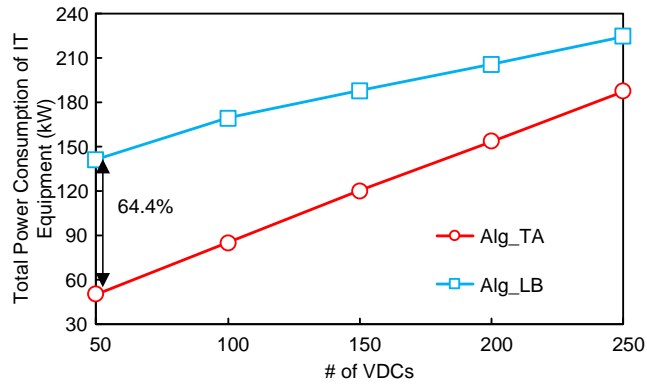

(b) Case B

Fig. 6. Total power consumption of IT equipment.

Fig. 6(b) shows the results of Case B. Again, we see that the temperature-aware scheme significantly outperforms the load-balanced scheme, by more than 64%, and the reason is the same as that explained for Case A.

*4) Trade-off between Maximum Rack Outlet Temperature and Total Power Consumption of IT Equipment*

As described in Section III, there is a trade-off between total power consumption of IT equipment and the maximum outlet temperature of racks. To show this trade-off, we solve the MILP model by changing the weight factor $\alpha$ in the objective function (2). Here we consider Case A for the test case, in which the number of VDCs is 20. Fig. 7 shows the results, where the results of the temperature-aware embedding algorithm are also plotted for comparison. Because the heuristic algorithm does not have the weight factor $\alpha$, we only show two lines for comparison. One corresponds to the maximum rack outlet temperature, and the other corresponds to the total power consumption of IT equipment.

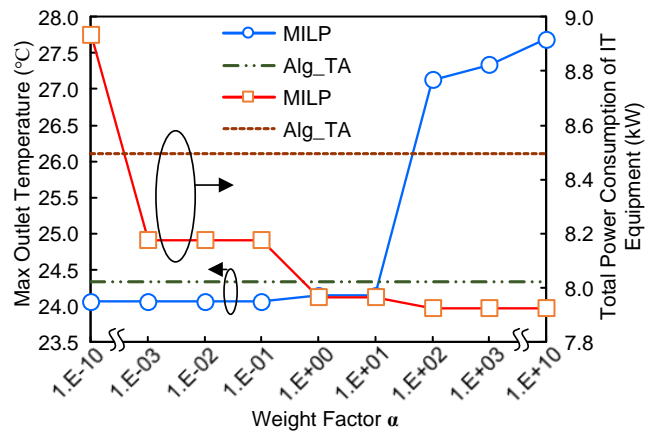

Fig. 7. Trade-off between maximum outlet temperature of racks and total power consumption of IT equipment under different weight factors $\alpha$.

25Based on the optimal results of the MILP model, we observe that, when $\alpha$ is smaller than 1, the maximum outlet temperature of racks rises very slowly with an increasing $\alpha$ and keeps almost constant at the very beginning. However, when $\alpha$ exceeds 1, the temperature value increases rapidly. In contrast, the total power consumption of IT equipment changes oppositely to reduce its value with an increasing $\alpha$. This is indeed in line with the function of a weight factor. A larger $\alpha$ means that minimizing the total power consumption of IT equipment is more important, and therefore, we see that the total power consumption of IT equipment becomes smaller. Moreover, it is interesting to see that the maximum outlet temperature of racks achieved by the temperature-aware embedding algorithm is sometimes lower than that obtained by the MILP model, which means that the proposed embedding scheme is very efficient to perform closely to the MILP model. Similar observations can be made for the total power consumption of all IT equipment. However, it seems that the total power consumption of IT equipment obtained by the algorithm is not as close to that of the MILP model compared with the maximum outlet temperature of racks. This is because the temperature-aware algorithm has put more emphasis on minimizing the maximum outlet temperature of racks, while considering the minimization of the total power consumption as the second-ranked objective.

## C. Results of Dynamic Scenario

For the dynamic scenario, following the approaches in most recent works of [10-14], we also assume that the arrivals of VDC requests follows a Poisson process with a rate of $\lambda$ VDC requests/hour, and the holding time of each VDC follows an exponential process with a mean time of $1/\mu$ hours. In this study, we set the mean duration $1/\mu$ to be 3 hours, and then evaluate the rejection ratio under different request arrival rates and temperature thresholds. We choose Case B for the case study, and a total of $10^6$ arrival VDC requests were simulated to calculate the VDC rejection ratio.

### 1) Rejection Ratio under Different VDC Arrival Rates

Fig. 8 shows the results of VDC rejection ratio with an increasing VDC arrival rate when the outlet temperature threshold is set to be 35°C. From the results, we observe that the temperature-aware scheme can significantly outperform the load-balanced scheme. This is because the temperature-aware scheme can well balance the outlet temperature of each physical rack when embedding VDCs. When the workload increases, the maximum outlet temperature of racks approaches the threshold quickly under the load-balanced scheme. Therefore, this will cause many VDCs rejected because of the violation of the temperature threshold. In contrast, the temperature-aware scheme can well balance the temperatures of all the racks and therefore can help a DC provision more VDCs without violating the temperature threshold.

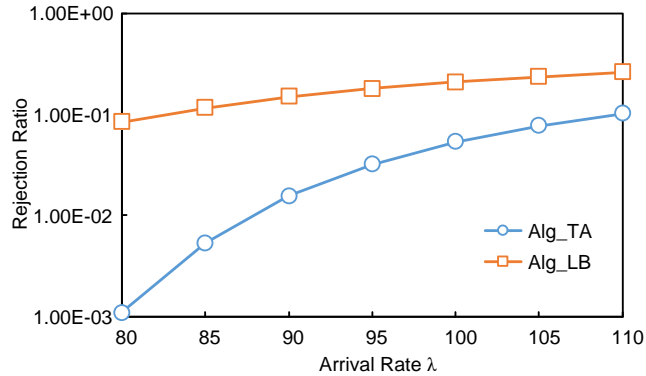

Fig. 8. Rejection ratio changes with increasing VDC arrival rate.

*2) Rejection Ratio under Different Outlet Temperature Thresholds*

We further evaluate how the VDC rejection ratio changes with different temperature thresholds. In this experiment, we set the arrival rate of VDC requests $\lambda$ to be 80 requests/hour. Fig. 9 shows the test results. From the results, we see that with an increasing temperature threshold, the rejection ratios obtained by both the temperature-aware scheme and the load-balanced scheme decrease. This is because a higher temperature threshold allows more VDCs to be embedded in a DC and therefore fewer VDC requests will be rejected. Also, it is observed that the temperature-aware embedding scheme always shows a much lower rejection ratio than that of the load-balanced scheme. This is again because the temperature-aware scheme can balance the outlet temperature of racks, thereby having fewer chances to violate the threshold when provisioning VDCs.

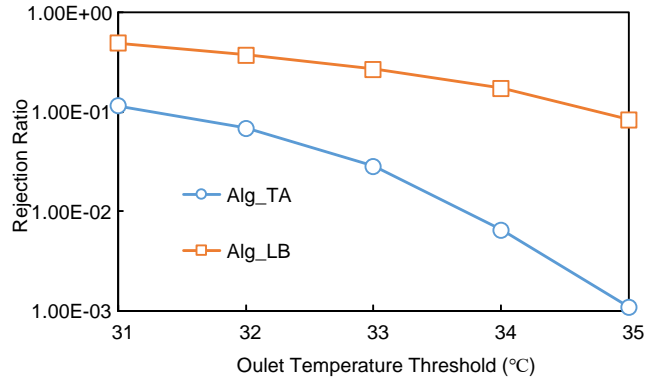

Fig. 9. Rejection ratio changes under different temperature thresholds.

*3) Traces of Maximum and Minimum Rack Outlet Temperatures*

We also show the real-time maximum and minimum outlet temperatures of racks under the dynamic VDC embedding scenario. Here we remove the constraint of temperature threshold and set the arrival rate of VDCs to be 80 VDC requests/hour.

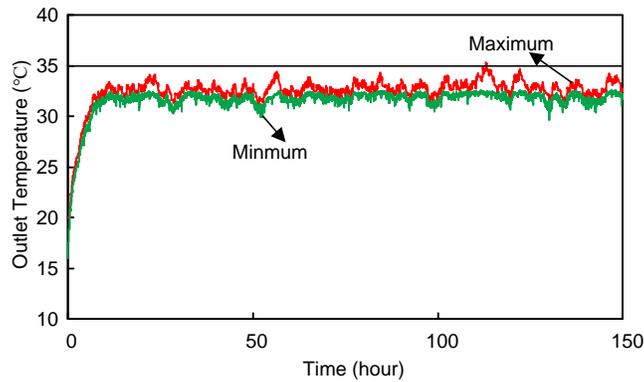

(a) Temperature-aware embedding.

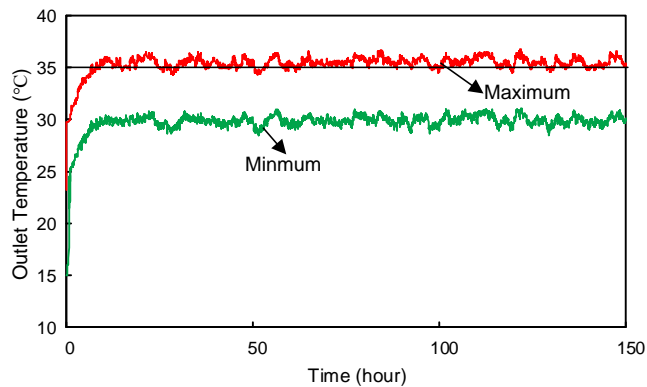

(b) Load-balanced embedding.

Fig. 10. Traces of maximum and minimum rack outlet temperatures.

Fig. 10(a) shows the traces of maximum and minimum rack outlet temperatures under the temperature-aware scheme and Fig. 10(b) shows the same traces under the load-balanced scheme. We can see that, under the temperature-aware scheme, the maximum and minimum outlet temperatures are close to each other with time going; in contrast, under the load-balanced scheme, there is a large gap between the maximum and minimum outlet temperatures, which is up to 5°C. Moreover, we see that there is a higher chance for the load-balanced scheme to have the maximum outlet temperature exceeding 35°C than the temperature-aware scheme. As such, if we set the temperature threshold to be 35°C, many VDC requests will be rejected under the load-balanced scheme, while almost none will be rejected under the temperature-aware scheme. The results again verify the capability of the temperature-aware scheme in balancing the rack outlet temperature even under dynamic VDC demands.

## VII. CONCLUSION

Heat generation in a DC can lead to many hot spots and therefore requires a cooling system in a DC to keep on running strongly to cool the whole DC environment. Hot spots affect not only the energy efficiency of a DC, but also the lifespans of IT equipment in the DC. To tackle this issue, we for the first time consider the approach to balance the temperature in a DC by properly embedding VDCs onto different racks. We develop an MILP model as well as an efficient temperature-aware VDC embedding algorithm. The results obtained under static VDC requests show that the proposed temperature-aware scheme can significantly reduce both the maximum outlet temperature of racks and the total power consumption of IT equipment compared to the load-balanced scheme. Also, the results obtained under dynamic VDC embedding show that the temperature-aware scheme has a significantly lower VDC rejection ratio than the load-balanced scheme under different rack temperature thresholds.

## REFERENCES

[1] WMware. [Online]. Available: http://www.vmware.com. [Accessed: 28-Oct-2019].
[2] Xen. [Online]. Available: http://xen.org. [Accessed: 28-Oct-2019].
[3] M. F. Bari, R. Boutaba, R. Esteves, L. Z. Granville, M. Podlesny, M. G. Rabbani, Q. Zhang, and M. F. Zhani, "Data center network virtualization: A survey," *IEEE Communications Surveys&Tutorials*, vol. 15, no. 2, pp. 909-928, second quarter, 2013.

[4] R. Mijumbi, J. Serrat, J.-L. Gorricho, N. Bouten, F. D. Turck, and R. Boutaba, "Network function virtualization: State-of-the-art and research challenges," *IEEE Communications Surveys&Tutorials*, vol. 18, no. 1, pp. 236-262, first quarter, 2016.
[5] D. Kreutz, F. M. V. Ramos, P. Verissimo, C. E. Rothenberg, S. Azodolmolky, and S. Uhlig, "Software-defined networking: A comprehensive survey," *Proceedings of IEEE*, vol. 103, no. 1, pp. 14-76, Jan. 2015.
[6] C. Guo, G. Lu, H. J. Wang, S. Yang, C. Kong, P. Sun, W. Wu, and Y. Zhang, "SecondNet: A data center network virtualization architecture with bandwidth guarantees," in Proc. *ACM CoNEXT* 2010.
[7] H. Ballani, P. Costa, T. Karagiannis, and A. Rowstron, "Towards predictable datacenter networks," in Proc. *SIGCOMM*, 2011, pp. 242-253.
[8] Amazon.com, Amazon elastic compute cloud (Amazon EC2). [Online]. Available: http://aws.amazon.com/ec2/. [Accessed: 28-Oct-2019].
[9] Microsoft.com, Microsoft Azure. [Online]. Available: http://azure.microsoft.com. [Accessed: 28- Oct- 2019].
[10] T. M. Nam, N. V. Huynh, L. Q. Dai, and N. H. Thanh, "An energy-aware embedding algorithm for virtual data centers," in Proc. *International Teletraffic Congress*, 2016, pp. 18-25.
[11] T. M. Nam, N. H. Thanh, H. T. Hieu, N. T. Manh, N. V. Huynh, and H. D. Tuan, "Joint network embedding and server consolidation for energy-efficient dynamic data center virtualization," *Computer Networks*, vol. 125, pp. 76-89, Jun. 2017.
[12] M. F. Zhani, Q. Zhang, G. Simon, and R. Boutaba, "VDC Planner: Dynamic migration-aware virtual data center embedding for clouds," in Proc. *IFIP/IEEE International Symposium on Integrated Network Management*, 2013, pp. 18-25.
[13] A. Amokrane, M. F. Zhani, R. Langar, R. Boutaba, and G. Pujolle, "Greenhead: Virtual data center embedding across distributed infrastructures," *IEEE Transactions on Cloud Computing*, vol. 1, issue 1, pp. 36–49, Jan.-Jun. 2013.
[14] A. Amokrane, R. Langar, M. F. Zhani, R. Boutaba, and G. Pujolle, "Greenslater: On satisfying green SLAs in distributed clouds," *IEEE Transactions on Network and Service Management*, vol. 12, no. 3, pp. 363-376, Sep. 2015.
[15] N. Le, "Data center top of rack and end of row design," [Online]. Available: https://lelunha.wordpress.com/2012/07/23/data-center-top-of-rack-and-end-of-row-design/. [Accessed: 28-Oct-2019].
[16] J. Choi, "Virtual machine placement algorithm for energy saving and reliability of servers in cloud data centers," *Journal of Network and Systems Management*, vol. 27, issue 1, pp. 149-165, Jan. 2019.
[17] A. Banerjee, T. Mukherjee, G. Varsamopoulos, and S. K. S. Gupta, "Cooling-aware and thermal-aware workload placement for green HPC data centers," in Proc. *International Conference on Green Computing*, 2010, pp. 245-256.
[18] ASHRAE TC 9.9, 2011 Thermal guidelines for data processing environments-Expanded data center classes and usage guidance, ASHARE, 2011.
[19] M. Dayarathna, Y. Wen, and R. Fan. "Data center energy consumption modeling: A survey," *IEEE Communications Surveys&Tutorials* vol. 18, no. 1, pp. 732-794, first quarter 2016.
[20] J. Moore, J. Chase, P. Ranganathan, and R. Sharma, "Making scheduling "cool": Temperature-aware workload placement in data centers," in Proc. *USENIX Annual Technical Conference,* 2005.
[21] Y. Yang, X. Chang, J. Liu, and L. Li, "Towards robust green virtual cloud data center provisioning," *IEEE Transactions on Cloud Computing*, vol. 5, no. 2, pp. 168-181, Apr.-Jun. 2017.
[22] Y. Han, J. Li, J.-Y. Chung, J.-H. Yoo, and J. W.-K. Hong, "SAVE: Energy-aware virtual data center embedding and traffic engineering using SDN," in Proc. *1st IEEE Conference on Network Softwarization*, 2015, pp. 1-9.
[23] Q. Zhang, M. F. Zhani, M. Jabri, and R. Boutaba, "Venice: Reliable virtual data center embedding in clouds," in Proc. *INFOCOM*, 2014, pp. 289-297.
[24] X. Wen, Y. Han, B. Yu, X. Chen, and Z. Xu, "Towards reliable virtual data center embedding in software defined networking," in Proc. *MILCOM*, 2016, pp. 1059-1064.
[25] G. Sun, Z. Xu, H. Yu, V. Chang, X. Du, and M. Guizani, "Toward SLAs guaranteed scalable VDC provisioning in cloud data centers," *IEEE Access*, vol. 7, pp. 80219-80232, Jun. 2019.
[26] H. Lo and W. Liao, "CALM: Survivable virtual data center allocation in cloud networks," *IEEE Transactions on Services Computing*, to appear. doi: 10.1109/TSC.2017.2777979.
[27] R. Yu, G. Xue, X. Zhang, and D. Li, "Survivable and bandwidth-guaranteed embedding of virtual clusters in cloud data centers," in Proc. *INFOCOM*, 2017, pp. 1-9.
[28] M. P. Gilesh, S. Satheesh, S. D. M. Kumar, and L. Jacob, "Selecting suitable virtual machine migrations for optimal provisioning of virtual data centers," in Proc. *ACM Symposium on Applied Computing*, 2018, pp. 22-32.
[29] M. P. Gilesh, S. D. M. Kumar, and L. Jacob, "Resource availability-aware adaptive provisioning of virtual data center networks," *International Journal of Network Management*, vol. 29, issue 2, Feb. 2019.
[30] F. Yan, T. T. Lee, and W. Hu, "Congestion-aware embedding of heterogeneous bandwidth virtual data centers with hose model abstraction," *IEEE/ACM Transactions on Networking*, vol. 25, no. 2, pp. 806-819, Apr. 2017.
[31] G. Sun, D. Liao, D. Zhao, Z. Xu, and H. Yu, "Live migration for multiple correlated virtual machines in cloud-based data centers," *IEEE Transactions on Services Computing*, vol. 11, issue 2, pp. 279-291, Mar.-Apr. 2018.

[32] X. Cao, I. Popescu, G. Chen, H. Guo, N. Yoshikane, T. Tsuritani, J. Wu, and I. Morita, "Optimal and dynamic virtual datacenter provisioning over metro-embedded datacenters with holistic SDN orchestration," *Optical Switching and Networking*, vol. 24, pp.1-11, Apr. 2017.

[33] W. Hou, Z. Ning, L. Guo, Z. Chen, and M. S. Obaidat, "Novel framework of risk-aware virtual network embedding in optical data center networks," *IEEE Systems Journal,* vol. 12, no. 3, pp. 2473–2482, Sep. 2018.

[34] Q. Tang, S. K. S. Gupta, and G. Varsamopoulos, "Energy-efficient thermal-aware task scheduling for homogeneous high-performance computing data centers: A cyber-physical approach," *IEEE Transactions on Parallel and Distributed Systems*, vol.19, no. 11, pp.1458-1472, Nov. 2008.

[35] X. Li, P. Garraghan, X. Jiang, Z. Wu, and J. Xu, "Holistic virtual machine scheduling in cloud datacenters towards minimizing total energy," *IEEE Transactions on Parallel and Distributed Systems*, vol. 29, issue. 6, pp. 1317-1331, Jun. 2018.

[36] S. Ilager, K. Ramamohanarao, and R. Buyya, "ETAS: Energy and thermal-aware dynamic virtual machine consolidation in cloud data center with proactive hotspot mitigation," *Concurrency and Computation: Practice and Experience*, vol. 31, issue 17, pp. 1-15, Sep. 2019.

[37] M. A. Islam, S. Ren, N. Pissinou, A. H. Mahmud, and A. V. Vasilakos, "Distributed temperature-aware resource management in virtualized data center," *Sustainable Computing: Informatics and Systems*, vol. 6, pp. 3-16, Jun. 2015.

[38] E. Pakbaznia, M. Ghasemazar, and M. Pedram, "Temperature-aware dynamic resource provisioning in a power-optimized datacenter," in *Proceedings of the Conference on Design, Automation and Test in Europe*, 2010, pp. 124-129.

[39] T. H. Cormen, C. E. Leiserson, R. L. Rivest, and C. Stein, *Introduction to algorithms*, 3rd editon, MIT press, 2009, ISBN: 978-0-262-03384-8.

[40] A. Greenberg *et al.*, "VL2: A scalable and flexlible data center network," in Proc. *ACM SIGCOMM Data Communication Conference*, 2009, pp. 51-62.

[41] Y. Ajiro and A. Tanaka, "Improving packing algorithms for server consolidation," *Proceeding of the Measurement Group's 2007 International Conference*, vol. 253, pp. 399-406, Dec. 2007.

[42] E. W. M. Wong and T.-S. Yum, ''Maximum free circuit routing in circuitswitched networks,'' in Proc. *INFOCOM*, 1990, pp. 934–937.

[43] L. Li, Y. Zhang, W. Chen, S. K. Bose. M. Zukerman, and G. Shen, "Naïve Bayes classifier-assisted least loaded routing for circuit-switched networks," *IEEE Access*, vol. 7, pp. 11854-11867, Jan. 2019.

[44] C. Guo, K. Xu, S. K. Bose, M. Zukerman, and G. Shen, "Efficient and green embedding of virtual data centers with mixture of unicast and multicast services," *IEEE Transactions on Cloud Computing*, to appear. doi: 10.1109/TCC.2019.2897273.

[45] C. Guo, G. Shen, Y. Yan, W. Chen, and S. K. Bose, "Consolidating virtual data centers with negative load correlations to alleviate performance degradation due to load burstiness (invited)," in Proc. *Asia Communications and Photonics Conference (ACP)*, 2018, pp. 1-3.